\documentclass[%
 reprint,
superscriptaddress,
 amsmath,amssymb,
 aps,
 prl,
]{revtex4-2}

\usepackage{graphicx}% Include figure files
\usepackage{dcolumn}% Align table columns on decimal point
\usepackage{bm}% bold math
\usepackage{physics}
\usepackage{nicematrix}
\usepackage{multirow} 
\usepackage{braket}
\usepackage{subfigure}   
\usepackage{floatrow}
\usepackage{verbatim}
\usepackage{color}
\usepackage[dvipsnames]{xcolor}
\usepackage{colortbl}
\usepackage[colorlinks,urlcolor=magenta,citecolor=blue,linkcolor= orange]{hyperref} 
\begin{document}

\preprint{APS/123-QED}

\title{Minute-Scale Photonic Quantum Memory}

\author{You-Cai Lv}
\author{Yu-Jia Zhu}
\author{Zong-Quan Zhou}
 \email{zq\_zhou@ustc.edu.cn}
\author{Chuan-Feng Li}
\author{Guang-Can Guo}

\affiliation{
Laboratory of Quantum Information, \\University of Science and Technology of China, Hefei, 230026, China
}
\affiliation{
Anhui Province Key Laboratory of Quantum Network, University of Science and Technology of China, Hefei 230026, China
}
\affiliation{
CAS Center For Excellence in Quantum Information and Quantum Physics,University of Science and Technology of China, Hefei, 230026, China
}
\affiliation{
Hefei National Laboratory, University of Science and Technology of China, Hefei 230088, China}
%\affiliation{These authors contributed equally}

\date{\today}

\bibliographystyle{naturemag}

\begin{abstract}
Long-lived storage of single photons is a fundamental requirement for enabling quantum communication and foundational tests of quantum physics over extended distances. While the implementation of a global-scale quantum network requires quantum storage times on the order of seconds to minutes, existing photonic quantum memories have so far been limited to subsecond lifetimes. Although $^{151}$Eu$^{3+}$:Y$_2$SiO$_5$ crystals exhibit substantially extended spin coherence times at the `magic' magnetic field, the concomitant weak optical absorption has until now prevented single-photon storage. Here, we overcome this challenge by integrating a noiseless photon echo protocol---which makes full use of the crystal's natural absorption for photonic storage---with a universally robust dynamical decoupling sequence incorporating adiabatic pulses to efficiently protect delocalized spin-wave excitation, enabling long-lived quantum storage at the `magic' magnetic field. At a storage time of 5.6 s, we achieve a time-bin qubit storage fidelity of 88.0 $\pm$ 2.1\%, surpassing the maximum fidelity attainable via classical strategies. Our device reaches a $1/e$ storage lifetime of 27.6 $\pm$ 0.5 s, enabling single-photon-level storage for 42 s with a signal-to-noise ratio greater than unity. This work establishes photonic quantum memory in the minute-scale regime, laying a solid foundation for global-scale quantum network and deep-space quantum experiments.
\end{abstract}

\maketitle

Quantum entanglement distribution is currently limited to 404 km in optical fibers \cite{Fiberlink1} and 1200 km via a single low-Earth-orbit satellite \cite{Satellite-based-entanglement}. Quantum repeaters \cite{QuantumRepeaters,Gisin} provide a scalable path beyond these ranges by dividing the total communication distance into shorter elementary links, synchronized via long-lived quantum memories. Realizing a global-scale quantum network will require quantum storage times on the order of seconds to minutes \cite{Gisin,900ms,lpx,Satellite}. Transportable quantum memories further broaden this vision by enabling worldwide quantum communication via the physical transport of stored quantum states, with target storage times approaching one hour \cite{Gundoǧan:24,double-slit}. Similarly, foundational experiments with deep-space quantum links---such as Bell tests over astronomical baselines and studies of entanglement dynamics in gravitational fields---also require photonic storage times lasting seconds to minutes \cite{QMforSpace,gravitation,long-baseline}.

However, existing photonic quantum memories remain limited to subsecond storage times \cite{100ms,20ms,pan22,pan458}. The current state of the art reaches a $1/e$ lifetime of approximately 0.1 s for absorptive-type quantum memories \cite{100ms,20ms} and 0.46 s for emissive-type systems \cite{pan458}, falling substantially short of the demands for global-scale quantum networks or advanced memory-assisted applications discussed above.

\begin{figure}[ht]
\includegraphics[width= 1.0\columnwidth]{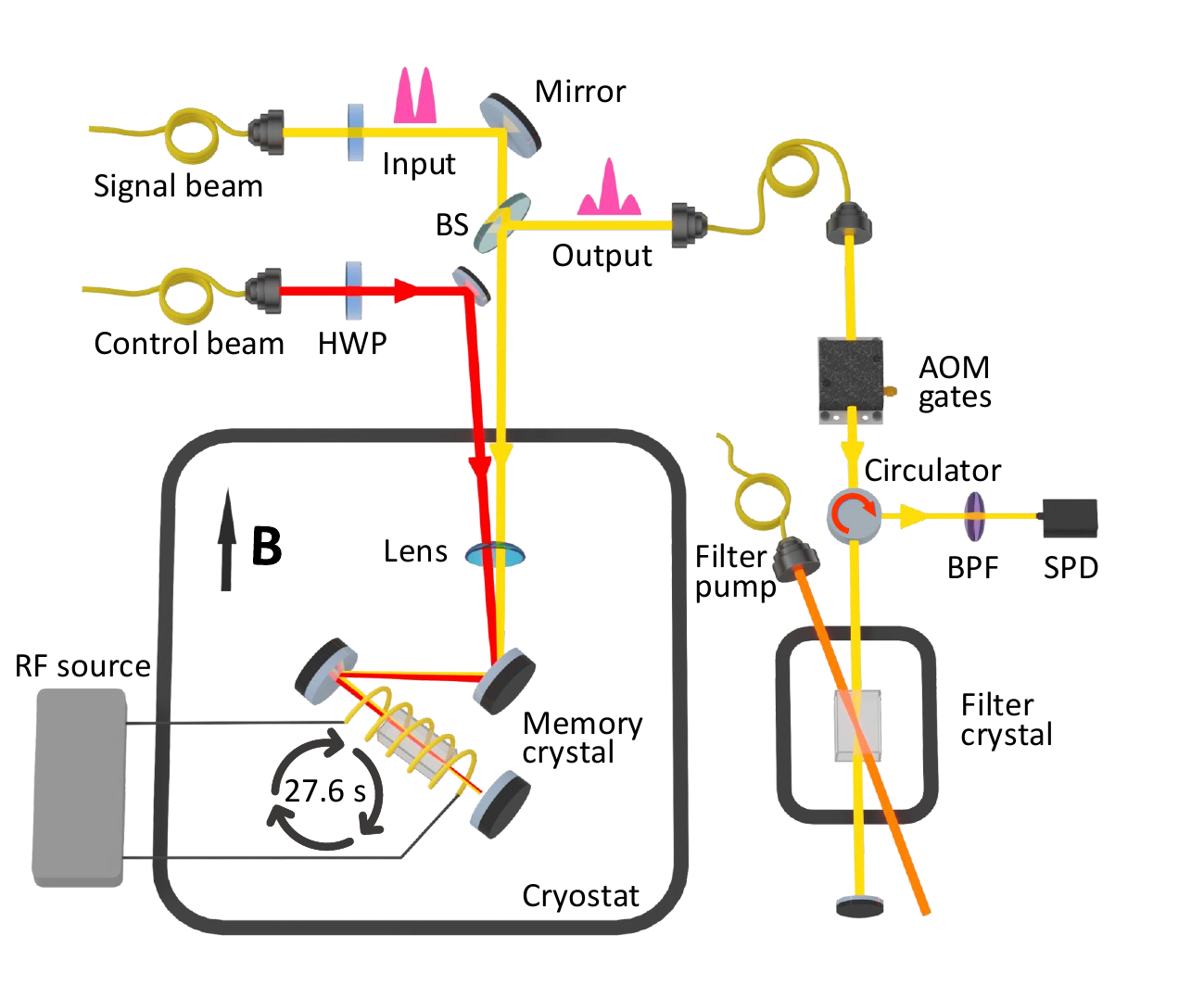}
\caption{\label{fig:setup}\textbf{Schematic of the experimental setup.} The polarization of signal and control beams are controlled with half-wave plates (HWPs), to be polarized parallel to the $D_1$ axis of the memory crystal (MC) to maximize interaction strength. Both beams overlap inside the MC an angle of 1.5$^{\circ}$. The signal beam propagates along the MC’s b-axis, guided by three steering mirrors, and is retro-reflected along the same path. A 95(R):5(T) beam splitter (BS) separates the output signal, which is then coupled into a single‑mode fiber. The MC is precisely aligned inside the magnet using two goniometer stages. Dynamical decoupling on the spin transitions is implemented by applying amplified RF pulses via solenoid coils surrounding the MC. Photonic time-bin encoded qubits are stored inside the MC with a $1/e$ lifetime of 27.6 s. The readout signal subsequently passes through a temporal gate based acousto-optic modulator (AOM), followed by spectral filtering with a double-passed filter crystal (FC) and a 0.5-nm band-pass filter (BPF). The FC is prepared by a pump beam to create a 0.8-MHz transparent spectral window centered at the signal frequency. The filtered signal is finally detected with a fiber‑coupled single‑photon detector (SPD). }
\end{figure}

Eu$^{3+}$:Y$_2$SiO$_5$ crystals are a leading candidate for long-lived photonic quantum memory \cite{6hour,mrr,zhangchao,20ms}, having demonstrated 1-hour storage of classical light at a `magic' magnetic field featuring zero first-order Zeeman (ZEFOZ) effects \cite{1hour}. Yet, the single-photon storage time remains confined to 0.1 s, achieved at near-zero field to maintain adequate optical absorption~\cite{20ms}. This limitation arises because operation at the ZEFOZ magnetic field, while advantageous for spin coherence, markedly diminishes optical absorption owing to the full removal of level degeneracy. Exacerbating this challenge is the difficulty of implementing long‑duration dynamical decoupling (DD) for a single delocalized spin-wave excitation. Consequently, the feasibility of long-lived quantum storage at the ZEFOZ magnetic field remains an open question~\cite{QM}.

In this work, we demonstrate a minute-scale photonic quantum memory by implementing the noiseless photon echo (NLPE) protocol \cite{NLPE} combined with a universally robust (UR) DD sequence \cite{UR4} in $^{151}$Eu$^{3+}$:Y$_2$SiO$_5$ crystals under a ZEFOZ magnetic field \cite{6hour}. The memory achieves a $1/e$ storage lifetime of 27.6 s—more than two orders of magnitude beyond previous records for absorptive photonic quantum memories \cite{100ms,20ms}. At a storage time of 5.6 s, it maintains an efficiency of 8.2\% and delivers a qubit storage fidelity of 88.0 $\pm$ 2.1\%, unambiguously surpassing the classical bound.

%and a natural absorption depth of 2 cm$^{-1}$ at `magic' magnetic field
Fig.~\ref{fig:setup} illustrates the experimental layout. The 580-nm laser is a frequency-doubled semiconductor laser (Toptica, TA-SHG) with a stabilized linewidth of approximately 1 kHz. Both the signal beam and the control beam are generated with double-passed acousto-optic modulators (AOMs). The control beam intersects the signal beam on the crystal at an angle of 1.5$^{\circ}$, with full width at half maximum (FWHM) of 250 \textmu m and 100 \textmu m for the control and signal beams, respectively. Both beams travel along the b-axis of the memory crystal (MC) with the polarization aligned to the $D_1$-axis. The MC is a 0.01\% $^{151}$Eu$^{3+}$:Y$_2$SiO$_5$ crystal with a size of $5\times4\times10$ mm$^{3}$ along the $D_1\times D_2\times b$. A solenoid coil with a diameter of 10 mm and a length of 18 mm surrounding the MC supplies the RF field for spin manipulation. The MC is cooled to 1.65 K using a low-vibration close-cycle cryostat (Attodry 2100). After the double-passed MC, the read-out pulse is temporally gated by AOMs and spectrally filtered by a double-passed filter crystal (FC) housed in a separate homemade cryostat. The FC is a 15-mm-long $^{151}$Eu$^{3+}$:Y$_2$SiO$_5$ crystal with a dopant concentration of 0.1\%. It is prepared with a 0.8-MHz-wide transparency window at the signal frequency, outside of which the absorption depth reaches 13.2.

In order to extend the spin coherence lifetime, the memory was operated at a ZEFOZ magnetic field of 1.2968 $\pm$ 0.0002 T, applied along the direction [-0.544, -0.601, 0.586] in the crystal's $D_1\times D_2\times b$ coordinate frame \cite{6hour,1hour}. We note that the reported field strength has now been calibrated with an NMR teslameter (Metrolab, PT2026), thus representing an accurate value. The orientation of the MC is finely adjusted using two goniometer stages \cite{1hour} to maximize the measured two-pulse spin coherence lifetime. After systematic optimization of the field magnitude and orientation, a two‑pulse spin‑echo coherence lifetime of $T_2=$ 18.7 $\pm$ 0.7 s (Fig.~\ref{fig:sequence}c) was achieved for the ZEFOZ transition between states $\ket3_g$ and $\ket4_g$. The two-pulse spin echo $T_2$ reported here is shorter than those achieved with shorter crystals \cite{6hour,1hour}, owing to the compromise in magnetic field homogeneity imposed by the 10-mm interaction length---necessary for efficient single-photon absorption---which in turn decreases the effective $T_2$ \cite{myz}. Employing a magnet with improved homogeneity would permit longer quantum storage times, as the two-pulse spin echo $T_2$ can be extended to approximately 50 s.

\begin{figure*}[ht]
\includegraphics[width= \textwidth]{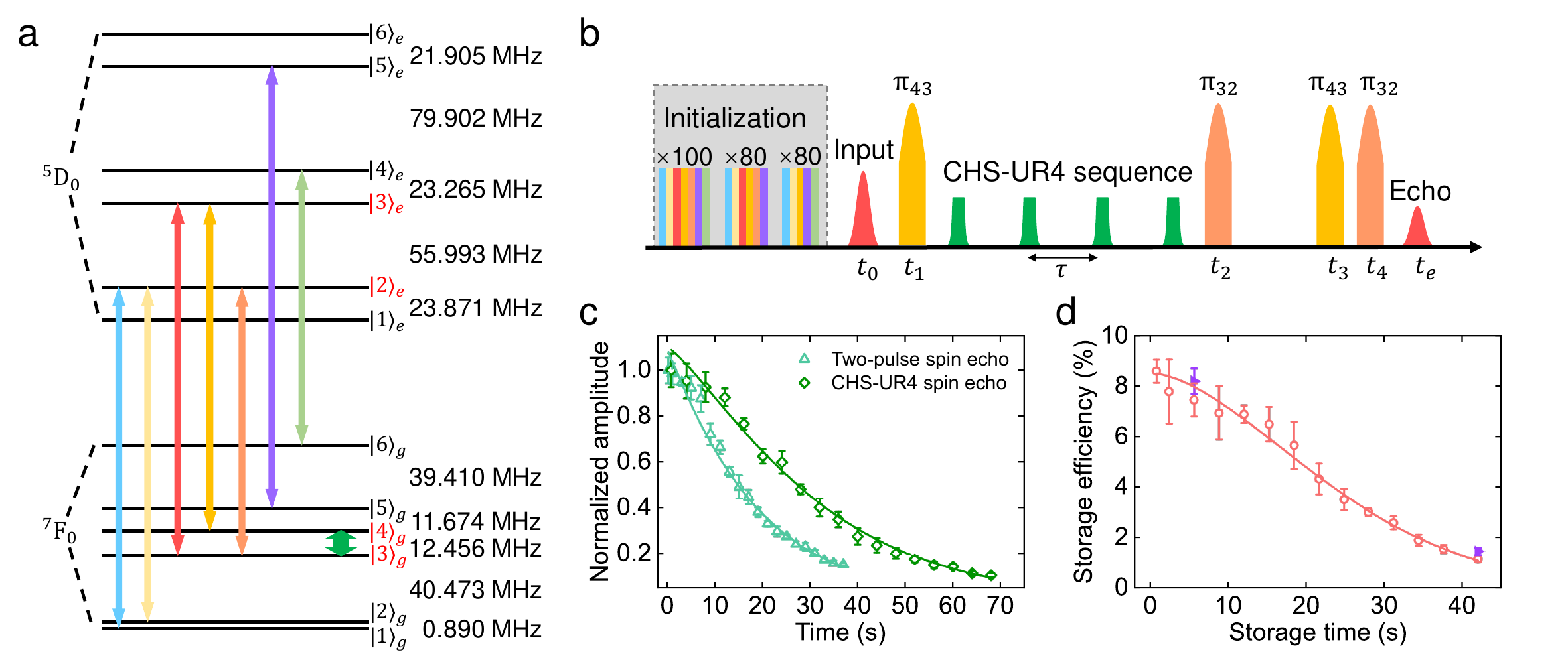}
\caption{\label{fig:sequence}\textbf{Characterization of the long-lived quantum memory.} \textbf{a}. Energy level structure of $^{151}$Eu$^{3+}$ in Y$_2$SiO$_5$ crystal under the ZEFOZ magnetic field of 1.2968 $\pm$ 0.0002 T. Colored lines represents optical and spin transitions: red for the signal beam, orange and coral for control beams; the green marker indicates the ZEFOZ spin transition at 12.456 MHz. \textbf{b}. Experimental sequence of the long-lived quantum memory. The gray region indicates initialization. This is followed by the NLPE-DD storage sequence, which employs a DD process based on the CHS-UR4 sequence. The final photonic echo is emitted at $t_e$ = $t_4$ + $t_3$ - $t_2$ - $t_1$ + $t_0$. \textbf{c}. Decay of spin echo amplitude with total evolution time for two-pulse spin echo (triangle) and the CHS-UR4 sequences (diamond). Spin echoes are detected via pulsed Raman heterodyne detection \cite{1hour}. The data are fitted to Mims' equation $e^{-(t/T_2)^m}$ \cite{Mims}, yielding $T_2$ = 18.7 s, $m$ = 1.05 for the two-pulse spin echo, and $T_2$ = 33.1 s, $m$ = 1.25 for the CHS-UR4 sequence. \textbf{d}. Photonic storage efficiency of the NLPE-DD memory as a function of storage time. The data are fitted to $e^{-(t/T_M)^m}$ with $T_M$ = 27.6 $\pm$ 0.5 s and $m$ = 1.70. Red circles denote measurements using classical light; purple triangles represent single-photon-level inputs. The slightly higher efficiency in single-photon measurements results from the uncorrected contribution from noise counts, which are negligible under classical inputs. All error bars indicate $\pm$1 standard deviation throughout.}
\end{figure*}

The application of a ZEFOZ magnetic field to $^{151}$Eu$^{3+}$:Y$_2$SiO$_5$ lifts all level degeneracies---including hyperfine doublets and magnetically inequivalent subsites---and consequently leads to a sharp reduction in optical absorption, measured here at approximately 0.5 cm$^{-1}$ for the double-passed MC. In this regime, standard quantum memory protocols that depend on spectral tailoring \cite{light–matter-interface,Efficient} to achieve atomic rephasing in an inhomogeneously broadened ensemble would suffer severe efficiency losses. We therefore adopt the NLPE protocol, which performs direct optical rephasing to make full use of the natural sample absorption for efficient quantum storage \cite{NLPE,jinming,tianxiang,lyp}.

\begin{table}[hb]
\renewcommand\arraystretch{1.2}
\caption{\label{BR}\textbf{Branching ratio of optical transitions.} Matrix representation of transition branching ratios between ground (columns) and excited (rows) states. Energy levels involved in the NLPE memory are highlighted in red.}
\begin{ruledtabular}
\begin{tabular}{ccccccc}
  &$\ket{1}_e$&\textcolor{red}{$\ket{2}_e$}&\textcolor{red}{$\ket{3}_e$}&$\ket{4}_e$&$\ket{5}_e$&$\ket{6}_e$ \\
\hline
$\ket{1}_g$& 0.57 & 0.18 & 0.12 & 0.10 & 0.01 & 0.02  \\
$\ket{2}_g$& 0.32 & 0.54 & 0.01 & 0.06 & 0.00 & 0.07  \\
\textcolor{red}{$\ket{3}_g$}& 0.06 & 0.18 & 0.44 & 0.10 & 0.01 & 0.21  \\
\textcolor{red}{$\ket{4}_g$}& 0.00 & 0.10 & 0.14 & 0.42 & 0.04 & 0.30  \\
$\ket{5}_g$& 0.04 & 0.00 & 0.23 & 0.27 & 0.08 & 0.38  \\
$\ket{6}_g$& 0.01 & 0.00 & 0.06 & 0.05 & 0.86 & 0.02  \\
\end{tabular}
\end{ruledtabular}
\end{table}

We first calculate the energy level structure of $^{151}$Eu$^{3+}$ in Y$_2$SiO$_5$ crystal under the ZEFOZ magnetic field, as illustrated in Fig.~\ref{fig:sequence}a. The optical transition probabilities are listed in Tab.~\ref{BR} which are calculated according to the hyperfine Hamiltonian of the ground and excited states provided in our previous work \cite{MA201832}. Implementation of NLPE protocol requires a four-level system whose two ground states are fixed to the ZEFOZ hyperfine transition $\ket3_g$ $\leftrightarrow$ $\ket4_g$. Among the optical transitions satisfying this constraint, the $\ket{3}_g$ $\leftrightarrow $ $\ket{3}_e$ line, which provides the highest transition strength, is selected as the signal transition. For the auxiliary excited state, we choose $\ket2_e$ because its transition to $\ket{3}_g$ is strong, while the $\ket{4}_g$ $\leftrightarrow $ $\ket{2}_e$  transition is weak; this asymmetry helps to suppress spontaneous-emission noise from the fully populated $\ket2_e$ level during the NLPE sequence \cite{NLPE}. 
%ChenDL

Fig.~\ref{fig:sequence}b sketches the time sequence of our long-lived photonic quantum memory. Initialization proceeds in three steps. First, a class-cleaning sequence \cite{NLPE,initialization} selects a class of ions within a 3 MHz bandwidth. Second, spin polarization \cite{NLPE,initialization} transfers all these ions to $\ket6_g$. Finally, five chirped pulses (0.8 MHz bandwidth) centered at $f_{12}$, $f_{22}$, $f_{43}$, $f_{55}$, and $f_{64}$ back-pump the ions into the target $\ket3_g$, where $f_{ij}$ represents the transition frequency for $\ket i_g$ $\leftrightarrow$ $\ket j_e$.
Following the initialization sequence, we obtain an absorption feature centered on the $\ket{3}_g$ $\leftrightarrow $ $\ket{3}_e$ transition, with a width of 0.8 MHz, a depth of 1.0, and located within a 3 MHz transparency window.

The input signal, centered at frequency $f_{33}$ and injected at time $t_0$, is a truncated Gaussian pulse with a total duration of 3 \textmu s and a FWHM of 1.5 \textmu s. The NLPE protocol requires four optical $\pi$ pulses to perform direct optical rephasing in the inhomogeneously broadened ensemble. To this end, we implement four adiabatic complex-hyperbolic-secant (CHS) pulses~\cite{CHS,NLPE}, which enable efficient and uniform $\pi$ rotations across the entire ensemble. The first $\pi_{43}$ pulse maps the optical signal onto the spin-wave excitation between states $\ket3_g$ $\leftrightarrow$ $\ket4_g$. A subsequent $\pi_{32}$ pulse converts this spin coherence into an optical coherence between $\ket2_e$ $\leftrightarrow$ $\ket4_g$. A second pair of $\pi_{43}$ and $\pi_{32}$ pulses then retrieves the optical echo signal at frequency $f_{33}$ \cite{tianxiang}. Since all population is eventually returned to the ground states and fluorescence from $\ket2_e$ is spectrally distinct from the signal, the final echo is emitted free from noise in principle. Here, $\pi_{mn}$ denotes a $\pi$ pulse driving the transition $\ket{m}_g$ $\leftrightarrow $ $\ket{n}_e$.

Extending the spin-wave storage lifetime for arbitrary photonic states requires a robust DD sequence that is insensitive to the initial state and immune to pulse errors. To this end, we implement a UR4 DD sequence \cite{UR4} and replace all rectangular $\pi$-pulses with adiabatic CHS RF pulses \cite{CHS,adiabaticpulse}, achieving optimized protection of the spin-wave quantum excitation. Furthermore, since the precise spacing between CHS pulses directly dictates the effective phase relationship—and thus the performance—of the UR4 sequence (see Supplementary Information for detailed analysis), we employ a rubidium frequency standard as the master time reference for pulse generation to ensure timing accuracy. This CHS-UR4 hybrid achieves optimal transfer efficiency while minimizing pulse-error-induced noise, obtaining a UR4 spin $T_2$ of 33.1 $\pm$ 1.6 s (Fig.~\ref{fig:sequence}c). Then we combine the NLPE memory with DD by applying the CHS‑UR4 sequence after the first optical $\pi_{43}$ pulse (Fig.~\ref{fig:sequence}b), achieving a $1/e$ quantum storage lifetime of $T_M$ = 27.6 $\pm$ 0.5 s for the NLPE-DD memory (Fig.~\ref{fig:sequence}d). To reduce integration time, most of the data are collected with classical light which agrees well with the single-photon data shown with purple triangles. We have observed that shorter DD pulse intervals caused notable RF-induced heating, degrading optical coherence and storage efficiency at short storage times. Consequently, the measured $T_M$ and the Mims factor $m$ \cite{Mims} differ from the values extracted from the CHS-UR4 spin echo fitting, with further discussion provided in the Supplementary Information.

\begin{figure}[ht]
%\begin{center}
\includegraphics [width= \textwidth]{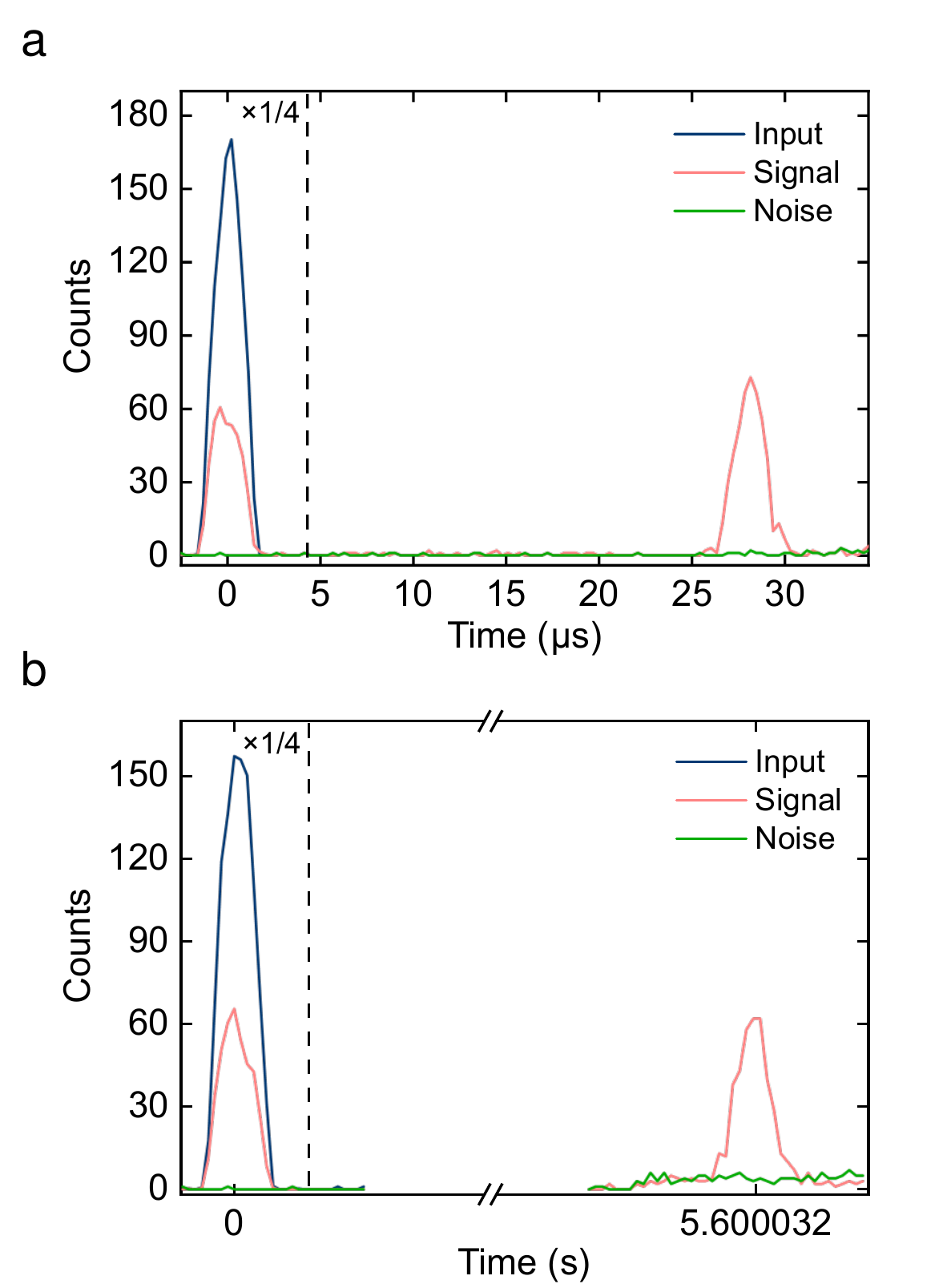}
\caption{\label{fig:SNR}\textbf{Storage of single-photon level inputs.} \textbf{a} and \textbf{b}. Photon counting histogram of NLPE memory (a) and NLPE-DD memory (b), measured using weak coherent input pulses with a mean photon number $\mu$ = 1.18 and integrated over 35,000 experimental trials. The blue, red, and green curves denote the input photon counts, output signal, and noise level (measured without input), respectively.  For clarity, data to the left of the dashed line are scaled down by a factor of 4. The shaded region indicates the 2.1 \textmu s detection window, yields SNRs of 65.2 $\pm$ 27.2 and 11.3 $\pm$ 2.5 at readout times of 28.1 \textmu s and 5.600032 s, respectively. }
%\end{center}
\end{figure} 

% This performance is comparable to that of NLPE memories operating at zero magnetic field~\cite{NLPE,tianxiang,lyp}.
Fig.~\ref{fig:SNR}a displays the photon counting histogram for the NLPE memory using weak coherent input pulses carrying a mean photon number of $\mu$ = 1.18 per pulse. Within a 2.1 \textmu s detection window,  the measured storage efficiency is 9.65 $\pm$ 0.5\% and a noise level of 0.17 $\pm$ 0.07\%, yielding a signal-to-noise ratio (SNR) of 65.2 $\pm$ 27.2 at a storage time of 28.1 \textmu s. Throughout this paper, the SNR are calculated using the noise-subtracted signal counts measured inside the detection window. Fig.~\ref{fig:SNR}b shows the corresponding histogram for the NLPE-DD memory under the same input conditions, using a DD pulse interval of $\tau$ = 1.4 s. At a readout time of 5.600032 s, the storage efficiency is 8.2 $\pm$ 0.5\%, with a SNR of 11.3 $\pm$ 2.5. Although the NLPE-DD memory maintains an efficiency comparable to that of the NLPE memory---reflecting the high rephasing efficiency of the CHS-UR4 sequence---the observed reduction in SNR stems from an increase in noise. This increased noise originates from imperfections in the CHS-UR4 sequence across the large ensemble, which leave residual population in the $\ket{4}_g$ after DD. This residual population is subsequently excited to $\ket{3}_e$ by the second $\pi_{43}$ pulse, leading to spontaneous emission noise. Despite this, the SNR achieved here is comparable to those of quantum memories incorporating DD at zero or low magnetic fields~\cite{20ms,lyp}, while extending the storage time by more than two orders of magnitude.

\begin{figure*}[ht]
%\begin{center}
\includegraphics [width= \textwidth]{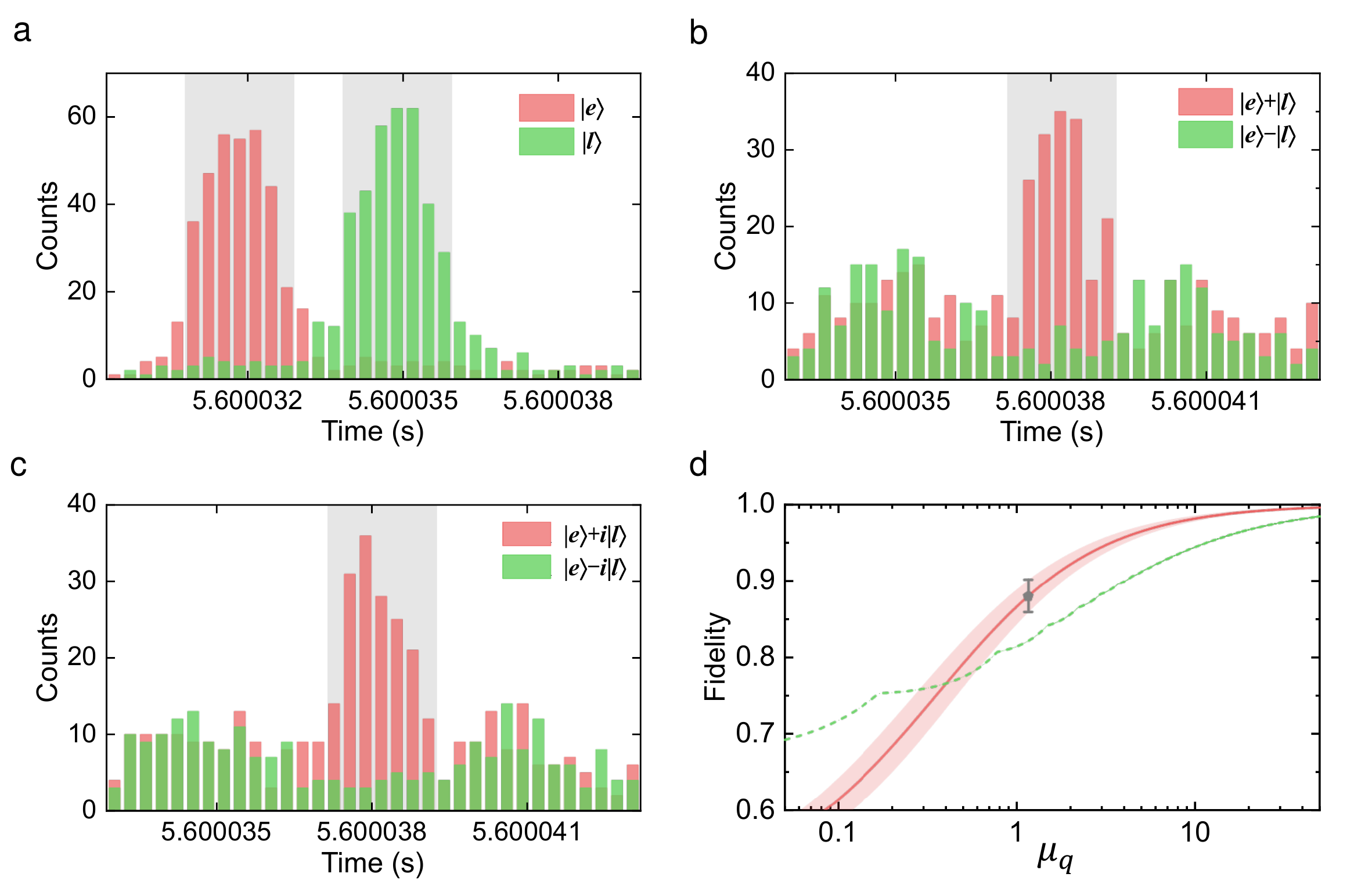}
\caption{\label{fig:timebin}\textbf{Storage of time-bin qubits for 5.600032 s.} \textbf{a}. Photon counting histogram for the input qubits $\ket{e}$ and $\ket{l}$. Output counts for $\ket{e}$ and $\ket{l}$ are shown as red and green bars, respectively. The gray-shaded region marks the 2.1 \textmu s detection window. The mean input photon number per qubit is $\mu_q$ = 1.16, integrated over 35,000 experimental trials. The measured fidelities are $F_{\ket{e}}$ = 92.7 $\pm$ 1.4\% and $F_{\ket{l}}$ = 92.7 $\pm$ 1.4\%. \textbf{b,c}. Photon counting histogram for the input qubit $\ket{e}+\ket{l}$ and $\ket{e}+i\ket{l}$, respectively. red and green bars correspond to constructive and destructive interference measurements for each qubit. All other experimental settings match those in \textbf{a}. The measured fidelities are $F_{\ket{e}+\ket{l}}$ = 85.8 $\pm$ 2.5\% and $F_{\ket{e}+i\ket{l}}$ = 85.6 $\pm$ 2.5\%. \textbf{d}. Classical fidelity limit for memory as a function of $\mu_q$. The green dashed line indicates the classical bound for a measure-and-prepare strategy under 8.2\% storage efficiency. The gray dot marks the measured memory fidelity of 88.0 $\pm$ 2.1\% at $\mu_q$ = 1.16. The solid red line represents the expected fidelity calculated from the experimentally determined efficiency and noise.}
\end{figure*}

To demonstrate that our device indeed works in the quantum regime, we encode the inputs with time-bin qubits \cite{NLPE,timebin}.
The four input qubit states are $\ket{e}$, $\ket{l}$, $\ket{e}+\ket{l}$ and $\ket{e}+i\ket{l}$, where $\ket{e}$ and $\ket{l}$ correspond to the early and late time bins, respectively. Each pulse retains identical parameters to those used in earlier tests, with a fixed separation of 3 \textmu s between $\ket{e}$ and $\ket{l}$. The eigen states $\ket{e}$ and $\ket{l}$ can be directly analyzed with standard NLPE-DD sequence. To analyze the superposition states, we split the second optical $\pi_{32}$ pulse into two $\pi/2$ pulses separated by 3 \textmu s, to read out the signal twice to effectively mimic the function of unbalanced Mach-Zehnder interferometer \cite{NLPE}. By tuning the phase between the two $\pi/2$ pulses, we modulate the interference of outputs to accomplish the analysis of the superposition states. The average photon number per qubit ($\mu_q$) is 1.16. Photon counting histograms for the four input qubits are presented in Fig.~\ref{fig:timebin} with a storage time of 5.600032 s. 

We denote the fidelity of input qubit $\ket{i}$ as $F_{\ket{i}}$. The total fidelity is computed as $F_t$ = $\frac{1}{3}\frac{F_{\ket{e}}+F_{\ket{l}}}{2}+\frac{2}{3}\frac{F_{\ket{e}+\ket{l}}+F_{\ket{e}+i\ket{l}}}{2}$. Here, the basis-state fidelities $F_{\ket{e}}$ and $F_{\ket{l}}$) are calculated as $\frac{S+N}{S+2N}$, where \textit{S} denotes the signal counts (noise excluded) and \textit{N} denotes the noise counts. Specifically for $F_{\ket{e}}$, $S+N$ and $N$ correspond, respectively, to the measured counts in the target time bin $\ket{e}$ and the counts in the orthogonal time bin $\ket{l}$. The superposition-state fidelities $F_{\ket{e}+\ket{l}}$ and $F_{\ket{e}+i\ket{l}}$ are evaluated as $\frac{V+1}{2}$, where $V = \frac{c_{max}-c_{min}}{c_{max}+c_{min}}$ is the interference visibility, with $c_{max}$ and $c_{min}$ representing the photon counts in the central time bin under constructive and destructive interference conditions, respectively. The measured storage fidelities are provided in Fig.~\ref{fig:timebin}, which give a $F_t$ of 88.0 $\pm$ 2.1\%. We calculate the maximal storage fidelity that can be achieved using classical prepare-and-measure strategy which takes into account the finite storage efficiency and the Poisson statistics of input coherent states \cite{single-atom,timebin}. The green dashed line in Fig.~\ref{fig:timebin}d represents this classical fidelity bound while the red solid line represents the expected storage fidelity of our device calculated according to the storage efficiency and measured noise. The measured fidelity $F_t$ at 5.600032 s violates the classical bound (82.1\%) by 2.8 standard deviations, unambiguously demonstrating this device operating in the true quantum regime. The cross over between these two lines indicates that our device operating beyond the classical limit with a minimal $\mu_q$ of 0.41.

\begin{figure}[ht]
%\begin{center}
\includegraphics [width= \columnwidth]{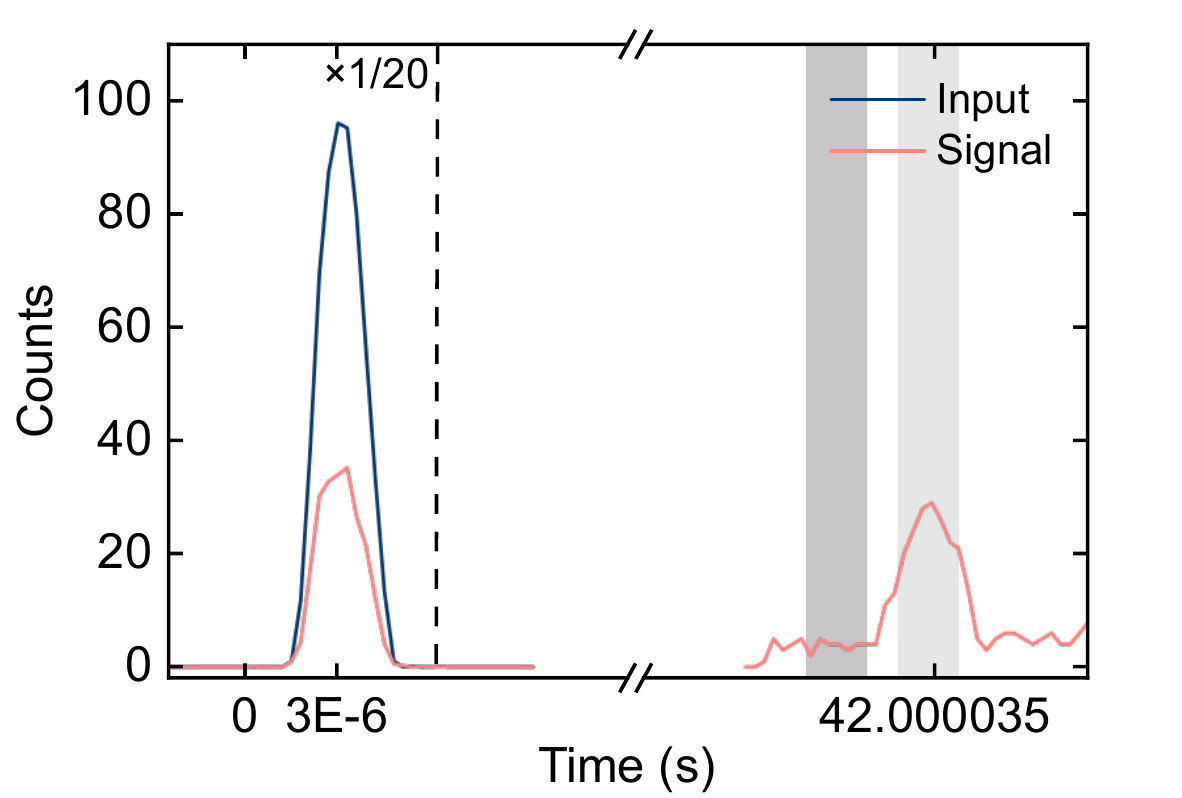}
\caption{\label{fig:42s}\textbf{Storage of qubit $\ket{l}$ for 42.00032 s.} Photon counting histogram of of the input qubit $\ket{l}$ encoded with weak coherent pulses with $\mu_q$ = 4.14, accumulated over 28,000 experimental trials. The blue curve indicates the input photon counts, and the red curve shows the storage process.  For visual clarity, photon counts to the left of the vertical dashed line are scaled down by a factor of 20. The dark and light shaded regions mark the detection windows for the orthogonal time bin $\ket{e}$ and target time bin $\ket{l}$, respectively.}
%\end{center}
\end{figure}

The fundamental fidelity limit for a quantum memory using true single-photon (Fock-state) inputs is 2/3~\cite{2/3}, equivalently requiring an SNR of 1:1. To probe the fundamental performance limits of our device, we extended the storage time to 42.000032 s by adjusting the CHS-UR4 pulse spacing to $\tau = 10.5\ \text{s}$ (Fig.~\ref{fig:42s}). To reduce integration time, measurements were conducted using an input qubit $\ket{l}$ with $\mu_q = 4.14$. The measured SNR of 5.54 $\pm$ 1.38 corresponds to a single-photon-equivalent SNR of 1.34 $\pm$ 0.33, which remains above the operational threshold for a quantum memory operating with ideal single-photon inputs. The reduction in SNR at longer storage times primarilly results from the decay of storage efficiency---which drops to 1.47\% at 42.000032 s (Fig.~\ref{fig:sequence}d)---while the noise floor remains nearly unchanged. This noise stability is achieved because the dominant noise source---residual population induced by the DD sequence---depends solely on the number of RF pulses applied, regardless of the interval between them. This behavior further confirms the successful minute-scale implementation of the UR4 DD sequence. Addressing the efficiency-noise trade-off in longer DD sequences will be pivotal to unlocking further gains in performance.

By implementing an NLPE-DD protocol in a $^{151}$Eu$^{3+}$:Y$_2$SiO$_5$ crystal at a ZEFOZ magnetic field, we have realized a photonic quantum memory with a $1/e$ lifetime of $T_M$ = 27.6 $\pm$ 0.5 s. The memory preserves photonic time-bin qubits beyond the classical fidelity limit after 5.6 s of storage---marking a 280-fold improvement over prior long-lived absorptive quantum memories \cite{20ms} and fulfilling a key requirement for global-scale quantum repeater networks \cite{900ms,lpx,Satellite} and fundamental physical tests with deep-space quantum links \cite{QMforSpace,gravitation}. 

Further extending the storage lifetime is essential for applications such as transportable quantum memories \cite{Gundoǧan:24,double-slit} and quantum communication across interplanetary or inter-spacecraft links \cite{long-baseline}. A storage time of several minutes can be straightforwardly achieved using a magnet with improved field homogeneity \cite{myz}. Substantially longer coherence can be attained by using $^{153}$Eu$^{3+}$ ions, which are expected to exhibit extended two-pulse spin echo $T_2$ due to their smaller magnetic $g$-tensor and reduced sensitivity to magnetic noise. Additional gains are feasible through the optimization of DD sequences with more pulses and the growth of higher-quality crystals \cite{myz}. Furthermore, near-unity storage efficiency could be reached by coupling the memory to an impedance-matched optical cavity \cite{mrr}, fully overcoming the weak-absorption constraint.

These advances position Eu$^{3+}$:Y$_2$SiO$_5$ crystals operating at ZEFOZ magnetic fields as the foundational hardware for large-scale quantum repeater networks whose transportable memories serve as mobile nodes, ushering in a flexible quantum architecture that mirrors today’s classical telecom infrastructure.

\bibliography{main,SM_main}% Produces the bibliography via BibTeX.

\bigskip
\noindent\textbf{Acknowledgments:} This work is supported by the National Natural Science Foundation of China (Nos. 12222411 and 11821404) and the Quantum Science and Technology-National Science and Technology Major Project (No. 2021ZD0301200). Z.-Q.Z acknowledges the support from the Youth Innovation Promotion Association CAS.\\
\noindent\textbf{Author contributions:} Z.-Q.Z. designed the experiment and supervised all aspects of this work; Y.-C.L. performed the experiment and analyzed the data with the help from Y.-J.Z.; Y.-C.L. and Z.-Q.Z. wrote the manuscript; Z.-Q.Z. and C.-F.L. supervised the project. All authors discussed the experimental procedures and results.  \\
\noindent\textbf{Competing interests:} The authors declare that they have no competing interests. \\
\noindent\textbf{Data and materials availability:} All data needed to evaluate the conclusions in the article are present in the paper or the supplementary materials.

\onecolumngrid
\newpage

\title{Supplementary Information for \\
Minute-Scale Photonic Quantum Memory}% Force line breaks with \\
\author{You-Cai Lv}
\author{Yu-Jia Zhu}
\author{Zong-Quan Zhou}
 \email{zq\_zhou@ustc.edu.cn}
\author{Chuan-Feng Li}
\author{Guang-Can Guo}

\affiliation{
Laboratory of Quantum Information, \\University of Science and Technology of China, Hefei, 230026, China
}
\affiliation{
Anhui Province Key Laboratory of Quantum Network, University of Science and Technology of China, Hefei 230026, China
}
\affiliation{
CAS Center For Excellence in Quantum Information and Quantum Physics,University of Science and Technology of China, Hefei, 230026, China
}
\affiliation{
Hefei National Laboratory, University of Science and Technology of China, Hefei 230088, China}
%\affiliation{These authors contributed equally}

\date{\today}

\bibliographystyle{naturemag}                        
\maketitle
%\renewcommand{\figurename}{Supplementary Fig.}
%\renewcommand{\thefigure}{\arabic{figure}}
%\listoffigures
%\captionsetup[figure]{name={Supplementary Fig.},justification=default}
\renewcommand{\figurename}{Fig.}
\renewcommand{\tablename}{Table.}

\setcounter{table}{0}
\renewcommand{\thetable}{S\arabic{table}}
\setcounter{figure}{0}
\renewcommand{\thefigure}{S\arabic{figure}}
\setcounter{equation}{0}
\renewcommand{\theequation}{S\arabic{equation}}

\newpage

\section{Supplementary Text}

\subsection{Experimental details}

\begin{figure*}[hb]
\includegraphics[width= 1.0\columnwidth]{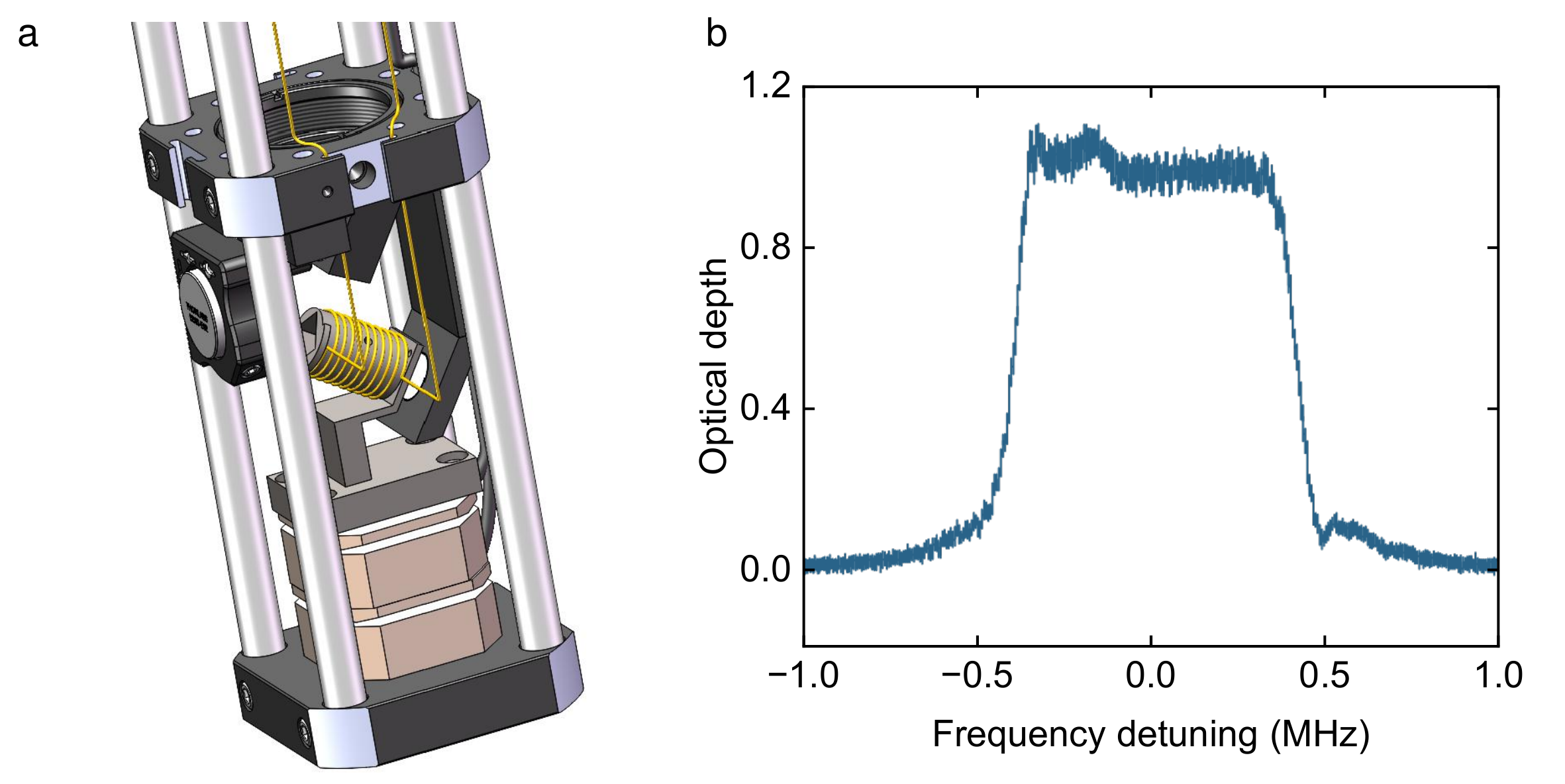}
\caption{\label{fig:SetupDetails}\textbf{Details about the sample space and the absorption structure of MC.} \textbf{a}. Three-dimensional simulated schematic of the sample space inside the MC-cryostat. \textbf{b}. The absorption profile of the MC after initialization process with zero detuning corresponding to the center of the $\ket{3}_g$ $\leftrightarrow $ $\ket{3}_e$ transition.}
\end{figure*}

Fig.~\ref{fig:SetupDetails}a shows the details about the sample space inside the memory-crystal (MC) cryostat. The MC is mounted in an acrylic sample holder mounted on two goniometers for magnetic-field alignment. The mirror placed in front of the MC with two-axis adjustment ensures the signal beam to propagate along the b-axis the MC. This optical configuration is crucial to ensure that the control light experience no birefringence, thereby enabling the application of ideal optical $\pi$ pulses for the MC in principle. A final mirror reflects the signal back through the MC, doubling the effective absorption length in a double-pass configuration. The control beam has a peak power of 320 mW at the crystal, reduced to 100 \textmu W during initialization. The signal beam operates at a peak power of 600 \textmu W and can be attenuated to the single-photon level using removable neutral density filters. The diameters of the control and signal beams at the center of the MC are approximately 250 \textmu m and 100 \textmu m. 

The incident signal power on the crystal is given by $\sqrt{W_1*W_2/0.95}$, where $W_1$ and $W_2$ denote the optical power measured before the cryostat and after the 95(R):5(T) beam splitter, respectively. This calibration procedure is consistently applied in both bright and single-photon-level measurements. A vibration sensor attached to the cryostat monitors the mechanical vibrations and generates a synchronization signal to initiate the experiment. This step ensures that the storage and readout process always occurs in a low-vibration window of the cryostat. Continuous operation of the storage and readout process can also be achieved by using dedicated vibration-isolation sample holders, as demonstrated in our previous work~\cite{zhangchao}. 
%From the front of the crystal to the single-photon detector (SPD), the optical path efficiency is 15.4\%, the detection efficiency of the SPD is 65\%, and the overall detection efficiency is 10.0\%.

Once the experiment begins, a chirped rectangular pulse centered at the signal frequency, with a bandwidth of 0.8 MHz, a duration of 1 ms, and a power of 60 \textmu W, is applied 200 times to the filtering crystal, generating a 0.8 MHz transparency window. The MC is then initialized through a sequence of class cleaning, spin polarization, and backburning steps \cite{initialization}. First, six chirped pulses (each with 3 MHz bandwidth, 1 ms duration) centered at frequencies $f_{12}$, $f_{22}$, $f_{33}$, $f_{43}$, $f_{55}$, and $f_{64}$ are applied to pump the six ground-state spin levels, where $f_{ij}$ represents the transition frequency for $\ket i_g$ $\leftrightarrow$ $\ket j_e$. Meanwhile, a chirped pulse centered at $f_{32}$ with the same parameters empties the frequency channel corresponding to the $\pi_{32}$ pulse. This sequence is repeated 100 times to selectively prepare a spectrally defined ion class. Subsequently, the pulse at $f_{64}$ is removed and the sequence is repeated 80 times to spin-polarize ions into state $\ket6_g$. Finally, inally, five chirped pulses (0.8 MHz bandwidth, 1 ms duration) at $f_{12}$, $f_{22}$, $f_{43}$, $f_{55}$, and $f_{64}$ are applied 80 times to back-pump the ions into the target state $\ket3_g$. In total, the initialization process for the MC takes a time of 1.580 s, which results in an isolated 0.8 MHz absorption feature on the $\ket{3}_g$ $\leftrightarrow $ $\ket{3}_e$ transition in the MC, as shown in Fig.~\ref{fig:SetupDetails}b. The four optical $\pi$-pulses for the NLPE protocol are implemented using adiabatic CHS pulses~\cite{CHS,NLPE}. The $\pi_{43}$ pulse has a duration of 4.1 \textmu s, a bandwidth of 0.8 MHz, and a peak power of 320 mW, while the $\pi_{32}$ pulse has a duration of 3.9 \textmu s, a bandwidth of 0.8 MHz, and a peak power of 320 mW.

\subsection{Characterization of NLPE protocol storage efficiency}

\begin{figure*}[hb]
\includegraphics[width= 1.0 \columnwidth]{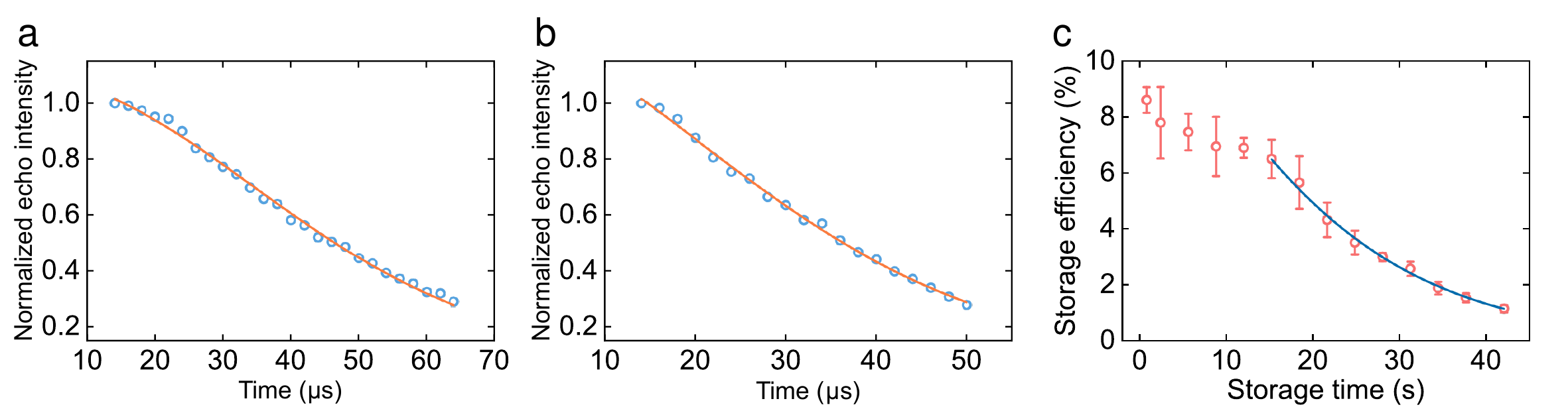}
\caption{\label{fig:NLPEdecay}\textbf{Efficiency decay of the NLPE memory.} \textbf{a, b}. NLPE echo intensity as a function of $t_{31}$ (a) and $t_{42}$ (b). The blue circles represent the measured values, and the orange curve represents the fit based on Eq.~\ref{Eq:gammafit}. \textbf{c}. The NLPE-DD memory efficiency versus storage times. Data after 15 s is fitted with Mims model $e^{-2(t/T^{\mathrm{spin}}_2)^m}$ \cite{Mims}, yielding $T^{\mathrm{spin}}_2$ = 36.3 $\pm$ 2.4 s and $m$ = 1.25.}
\end{figure*}

The storage efficiency of the NLPE memory can be modeled as~\cite{tianxiang}:
\begin{eqnarray}
\label{Eq:gammafit}
\eta = d^2e^{-d}(\eta_{\mathrm{control}})^4e^{-\frac{{\Gamma_{34}}^2{t_{31}}^2}{2\ln{2}/\pi^2}}e^{-\frac{{\Gamma_{\bar{2}\bar{3}}}^2{t_{42}}^2}{2\ln{2}/\pi^2}-2\gamma t_{42}},
\end{eqnarray}
where $d$ denotes the effective absorption depth after initialization, $\eta_{\mathrm{control}}$ represents the transfer efficiency of each optical $\pi$-pulses, $t_\mathrm{kl}$ represents the time interval between the $\mathrm{k}$-th $\pi$-pulse and the $\mathrm{l}$-th $\pi$-pulse., $\Gamma_{34}$ is the inhomogeneous broadening of the $\ket{3}_g$ $\leftrightarrow $ $\ket{4}_g$ transition, $\Gamma_{\bar{2}\bar{3}}$ is the inhomogeneous broadening of the $\ket{2}_e$ $\leftrightarrow $ $\ket{3}_e$ transition, and $\gamma$ is the effective optical decoherence rate.
We measure the decay of the NLPE echo as a function of $t_{31}$ and $t_{42}$ (Fig.~\ref{fig:NLPEdecay}), and obtain the fitting results using Eq.~\ref{Eq:gammafit}: $\Gamma_{34}$ = 7.7 $\pm$ 0.1 kHz, $\Gamma_{\bar{2}\bar{3}}$ = 8.4 $\pm$ 0.5 kHz, $\gamma$ = 5.9 $\pm$ 1.8 kHz. After substituting the measured $\eta$ and $d$, we obtain 
$\eta_{\mathrm{control}}$ = 82\% for the transfer efficiency of the optical $\pi$-pulses. The observed reduction in transfer efficiency, compared to the zero-field NLPE memory \cite{NLPE,tianxiang}, is due to the vertically reflected double-pass configuration used here, where the signal beam is focused onto the mirror rather than at the center of the MC. Consequently, the beam diameter increases within the crystal, degrading spatial overlap with the control beam. This can be mitigated in future designs by using a dual-window cryostat, or by adopting an integrated waveguide architecture \cite{tianxiang}.

In the NLPE-DD memory, when the DD pulse intervals are short, RF-induced heating becomes significant, leading to degradation of the optical storage efficiency at short storage times. As a result, only the $1/e$ storage lifetime can be extracted from the efficiency decay curve (as shown in Fig. 2d of the main text), and the spin-coherence time $T^{\mathrm{spin}}_2$ cannot be reliably determined through fitting of optical storage efficiency. To mitigate the influence of RF heating, we perform fitting only on data points with storage times longer than 15 s (Fig.~\ref{fig:NLPEdecay}c), where heating effects are no longer significant. The fit yields a spin-coherence time of $T^{\mathrm{spin}}_2 = 36.3 \pm 2.4$ s with a Mims factor of $m = 1.25$, which is consistent with the CHS-UR4 spin-echo $T_2$ reported in the main text.

\bigskip
\bigskip
\bigskip
\bigskip

\subsection{The CHS-UR4 sequence}

The performance of DD over an ensemble-based quantum memory is fundamentally determined by the quality of the individual pulses, the noise-suppression capability of the sequence, and its robustness to pulse errors. In practice, the dominant limitation of DD is pulse imperfections, often exceeding the impact of environmental noise~\cite{RobustDD}. This is especially true for our large, spectrally broadened atomic ensemble. The UR sequences achieve highly robust DD by engineering the pulse phases so that pulse-error contributions cancel at the sequence level, while offering arbitrary-order compensation against slowly changing environmental dephasing. In principle, they can compensate pulse errors originating from any experimental parameter and remain compatible with arbitrary pulse shapes \cite{UR4}. To minimize noise associated with a large number of pulses, here we employ the simplest variant, the UR4 sequence. For rectangular $\pi$-pulses, the phases of the UR4 sequence are given by 0, $\phi_2$, $\pi+2\phi_2$, and $3\pi+3\phi_2$, where $\phi_2$ is an arbitrary variable.

In practice, the performance of rectangular $\pi$-pulses is limited by their narrow bandwidth and sensitivity to pulse errors, particularly given the weak nuclear spin transition and limited RF power available. We therefore replace them with more robust adiabatic CHS pulses~\cite{CHS,adiabaticpulse}. However, because the phase of a CHS pulse is defined differently from that of a rectangular $\pi$-pulse, further clarification is necessary. The CHS pulse is described by:
\begin{eqnarray}
\label{Eq:CHS}
\Omega_0 \mathrm{sech}(\beta t) \cos[\omega_c t - \mu \ln(\mathrm{sech}(\beta t)) + \phi_0],
\end{eqnarray}
where $\Omega_0$ is the maximum Rabi frequency, $\beta$ characterizes the FWHM of the pulse, $\mu$ is the chirp strength parameter, $\phi_0$ denotes the constant phase, and \textit{t} ranges from $-\frac{T}{2}$ to $\frac{T}{2}$, with \textit{T} representing the pulse duration. From the Eq.~\ref{Eq:CHS}, the CHS pulse has an amplitude envelope $\Omega(t) = \Omega_0\mathrm{sech}(\beta t)$ and a dynamic phase $\phi(t) = - \mu \ln(\mathrm{sech}(\beta t)) + \phi_0$.
Previous studies indicate that, under adiabatic conditions, an adiabatic rapid passage (ARP) produces the same effect as a $\pi$-rotation of the Bloch vector about an axis located in the xy-plane of the Bloch sphere \cite{adiabaticpulse}. The phase of the rotation axis can be written as:
\begin{eqnarray}
\label{Eq:ARPphase}
\phi^{\mathrm{ARP}}_{0} = \frac{\pi}{2} - \frac{1}{2}\int_{-\frac{T}{2}}^{\frac{T}{2}} \sqrt{\Omega(t)^2 + \Delta(t)^2} \, dt + \frac{1}{2}[\phi(-\frac{T}{2}) + \phi(\frac{T}{2})],
\end{eqnarray}
where \textit{T} is the pulse duration, $\Delta(t) = -\dot{\phi(t)}$ (assuming zero frequency detuning at the center of the ARP), $\phi(-\frac{T}{2})$ and $\phi(\frac{T}{2})$ denote the pulse phases at times \textit{t} = $-\frac{T}{2}$ and \textit{t} = $\frac{T}{2}$, respectively. Substituting the CHS pulse parameters into Eq.~\ref{Eq:ARPphase} yields the rotation-axis phase for the CHS pulse:
\begin{eqnarray}
\label{Eq:CHSphase}
\phi^{\mathrm{CHS}}_{0} = \frac{\pi}{2} - \frac{1}{2}\int_{-\frac{T}{2}}^{\frac{T}{2}} \sqrt{(\Omega_0\mathrm{sech}(\beta t))^2 + (\mu \beta \mathrm{tanh}(\beta t))^2} \, dt - \mu \ln(\mathrm{sech}(\beta \frac{T}{2})) + \phi_0,
\end{eqnarray}
Thus, when all other CHS pulse parameters are held constant, the phase difference between CHS pulses depends solely on $\phi_0$, which aligns with the behavior of rectangular $\pi$-pulses.

For the CHS-UR4 sequence, setting $\phi_2 = \pi/2 + \delta$ results in the four pulse phases: 0, $\pi/2+ \delta$, $2\delta$, and $\pi/2 + 3\delta$, where $\delta$ ranges from 0 to $2\pi$. Notably, the UR4 sequence reduces to the XY4 sequence when $\delta = 0$~\cite{XY4}.  The CHS-UR4 sequence features a linearly increasing phase $N \cdot \delta$, where $N$ is the pulse index. Therefore, while the CHS-UR4 sequence uses a similar configuration as the XY4 sequence, it relaxes the requirement for strict $\pi/2$ phase shifts between adjacent pulses. Accordingly, by precisely controlling the time intervals between CHS pulses, which ensures the linearly increasing phase, and embedding the phase pattern [0, $\pi/2$, 0, $\pi/2$] within each CHS pulse waveform, the CHS-UR4 sequence can be accurately generated using standard arbitrary waveform generators (AWGs).

We employ an AWG (Zurich Instruments, HDAWG) to generate the CHS pulse waveforms. Given that the DD pulse intervals are on the order of seconds or even tens of seconds, the internal crystal-oscillator clock of the HDAWG cannot maintain sufficiently stable intervals (and the corresponding phase relationships) between CHS pulses. To ensure temporal stability across the DD sequence, we use a rubidium frequency standard as an external clock for the AWG. Each RF CHS pulse in our experiments has a duration of 3.9 ms, a bandwidth of 22 kHz, and a peak power of 150 W.

In summary, by combining the robustness of the UR4 sequence with that of individual CHS pulses, the CHS-UR4 DD sequence emerges as an excellent candidate for high-performance ensemble-based photonic quantum memories. For future work, efforts should focus on balancing the storage efficiency of the NLPE-DD memory at high pulse numbers against the accumulated population error induced by multiple pulses, thereby maximizing the achievable quantum memory lifetime.

%\bibliography{main,SM_main}% Produces the bibliography via BibTeX.

\end{document}